\documentclass[aps, pra, reprint, superscriptaddress,longbibliography]{revtex4-2}

\usepackage{lineno,hyperref}
\usepackage{eurosym}
\usepackage{amsmath,amssymb,graphicx}
\usepackage{color}

\modulolinenumbers[5]
\newcommand{\bitem}{\begin{itemize}}
\newcommand{\fitem}{\end{itemize}}
\newcommand{\beq}{\begin{equation}}
\newcommand{\eeq}{\end{equation}}
\newcommand{\beqa}{\begin{eqnarray}}
\newcommand{\eeqa}{\end{eqnarray}}

\begin{document}

\title{Dissipative optomechanical preparation of non-Gaussian mechanical entanglement}

\author{G. D. de Moraes Neto}
\email{gdmneto@zjnu.edu.cn}
\affiliation{Department of Physics, Zhejiang Normal University, Jinhua 321004, China}

\author{V. Montenegro}
\email{vmontenegro@uestc.edu.cn}
\affiliation{Institute of Fundamental and Frontier Sciences, University of Electronic Science and Technology of China, Chengdu 610051, China}

\begin{abstract}
Entanglement had played a crucial role in developing frontier technologies as a critical resource, for instance, in quantum teleportation and quantum sensing schemes. Notably, thanks to the ability to cool down the vibrational modes of mechanical oscillators to its quantum regime, entanglement between mechanical modes and the production of nonclassical mechanical states have emerged as central resources for quantum technological applications. Thus, proposing deterministic schemes to achieve those tasks is of paramount importance. While the dominant scheme for bipartite mechanical entanglement involves Gaussian optomechanical interactions (linearized regime) to generate two-mode squeezed vacuum states, entangling two-modes exploiting the bare non-Gaussian optomechanical interaction (nonlinear strong single-photon regime) remains less covered. This work proposes an on-demand scheme to engineer phononic non-Gaussian bipartite entanglement in the nonlinear regime by exploiting cavity dissipation. Interestingly, our protocol (operating in the resolved sideband and photon blockade regime) renders the possibility of achieving a high degree of steady-state entanglement. We further show that our deterministic scheme is robust in the presence of decoherence and temperature within state-of-the-art optomechanics, along with the required conditions to obtain non-Gaussianity of the achieved bipartite mechanical steady-state.
\end{abstract}

\maketitle

\section{Introduction}
Generating deterministic quantum resources is at the heart of quantum technologies~\cite{QT, Res}. Within quantum resource theories, entanglement is one of the most celebrated resources as it provides unprecedented advantages when the system’s dynamics are limited to local quantum operations and classical communication~\cite{Res}. Highly entangled states in the fabric of a quantum system have proven to power many quantum information tasks~\cite{EE}, such as quantum teleportation protocols~\cite{tele} and their usage for quantum sensing applications~\cite{sensing}. Notable advancements on the entanglement of Gaussian states have been put forward, including the generation of two-mode squeezed vacuum states~\cite{exp1, gnl3, Tan2013, Abdi2015, Wang2013} and the exceptional scalability across up to one million entangled optical modes~\cite{vm}. While Gaussian states are freely available in the laboratory and accompanied by a robust theoretical framework~\cite{G1,G2}, it has been shown that non-Gaussian features are intrinsically a quantum resource for numerous quantum protocols~\cite{NG_1,WLN} and are essential to attain a quantum computational advantage~\cite{NG_2}. Hence, it is highly desirable to generate an on-demand non-Gaussian entangled state for practical quantum applications.

Optomechanical systems naturally emerge as excellent platforms for generating non-Gaussian states due to the intrinsic non-Gaussian light-matter interaction in those systems~\cite{Optomechrev,PLAreview, chinesreview, cklaw-hamiltonian}. Within the nonlinear (non-Gaussian) interaction regime, numerous proposals for the quantum information field include entanglement distillation between optical modes~\cite{Montenegro2019}, preparation of massive quantum superpositions~\cite{Kleckner2008,massive}, production of nonclassical states for the light~\cite{Bose1997, Mancini1997} and the mechanics~\cite{NooN,PRA.2013.88.063819,PRA.2013.87.053849,Phono_Fock}, and quantum optomechanical sensors~\cite{Montenegro2020, sofia-gravimetre}. Nonetheless, accessing the strong nonlinear single-photon regime is still experimentally demanding. To overcome this obstacle, proposals to strengthen the light-matter coupling include the squeezing of the cavity mode~\cite{gexp1, Li2015, Vitali2007}, the enhancement of the coupling by the mediation of a qubit system~\cite{Pirkkalainen2015, Aporvari2021}, the usage of parametric driving field applied to the mechanical resonator~\cite{gexp2}, or typically through strong driving of the cavity mode~\cite{Optomechrev}. Although driving the cavity mode leads to exciting quantum information schemes such as quantum state transfer in optomechanical arrays~\cite{QST} and the generation of deterministic mechanical entanglement~\cite{exp2, exp3, Hartmann2008, Genes2008} and probabilistic~\cite{Pirandola2006, Borkje2011, Abdi2012} schemes, the strong external pumping linearizes the optomechanical interaction. Thus, the non-Gaussian feature from the nonlinear optomechanical interaction vanishes, and consequently, one cannot generate non-Gaussian entangled states of either the optical or the vibrational modes in the linearized regime from the dynamics alone. Nevertheless, one can use the linear interaction followed by measurements to obtain non-Gaussian states, as demonstrated in recent experiments of  phonon subtraction/addition~\cite{exp1-add} and single excitation mechanical entangled states~\cite{exp2-remote}.

A second challenge for the generation of non-Gaussian entangled states in the nonlinear optomechanical regime is the problem of decoherence. In this direction, efficient techniques for preparing and protecting the system's relevant degrees of freedom from quantum noise~\cite{Suter2016, Duan1997} include decoherence-free subspaces~\cite{DFS, destorsz}, dynamical decoupling~\cite{DD}, and reservoir engineering~\cite{PCZ, RE, PNS} techniques. Even though each of the above proposals has its merits for bypassing decoherence, quantum reservoir engineering has attracted much attention due to the ample flexibility in building the needed decoherence channels. In quantum reservoir engineering schemes, one builds non-unitary dynamics via a set of dissipative operators. The interplay between the natural decoherence channels and the engineered reservoir leads deterministically to a desired target steady-state. This scheme, independent of the system's initial state, has been employed for dissipative preparations of many-body quantum states~\cite{Many,Ne1,Ne2}, the superposition between two wave packets~\cite{SP,SP1}, universal dissipative quantum computation~\cite{UniDissQC}, and analog quantum simulation in open systems~\cite{ExpDissEng}. Furthermore, already been experimentally realized in trapped ions~\cite{IT}.

This work presents a deterministic protocol for generating non-Gaussian mechanical entanglement by exploiting the bare non-Gaussian nature of the optomechanical interaction. Our dissipative protocol considers a multimode optomechanical system in the strong single-photon optomechanical regime~\cite{chinesreview}, where two mechanical oscillators are placed inside a driven two-mode optical resonator. Without linearizing the optomechanical system, we present a general description of the normal modes of the mechanical oscillators within the sideband~\cite{Optomechrev} and the photon-blockade regime~\cite{blockade}. Interestingly, we show that our protocol allows the possibility of achieving a high degree of steady-state entanglement. We also show that our scheme is robust against the effects of decoherence and temperature. Finally, we study the conditions to obtain non-Gaussianity of the generated mechanical steady-state.

The rest of the paper is structured as follows: In Sec.~\ref{sec:the-model}
we derive an effective Hamiltonian in the blue-sideband and photon blockade regime for the normal modes of the mechanical oscillators. In addition, we present the dissipative mechanism inspired by the reservoir engineered techniques. In Sec.~\ref{sec:results}, we outline our results, including (i) the analysis of the steady-state entanglement and purity of the generated mechanical steady-state; (ii) the impact of decoherence and temperature in our scheme; and (iii) the degree of non-Gaussianity of the bimodal mechanical target state. In Sec.~\ref{sec:experimental-feasibility}, we present the experimental feasibility of our protocol. Finally, we present the concluding remarks in Sec.~\ref{sec:conclusions}.

\section{The Model}

\label{sec:the-model} 
\begin{figure}[t]
	\centering \includegraphics[width=1\linewidth]{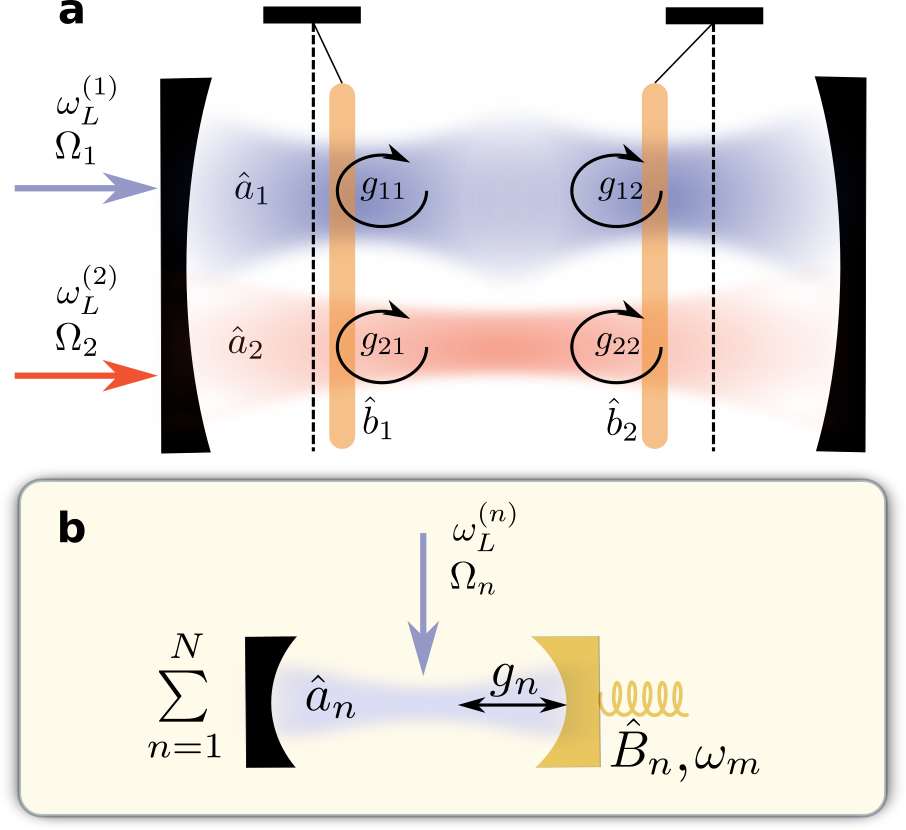}
	\caption{\textbf{a}  Schematic of two vibrating membranes carefully placed inside a two-mode cavity. The cavity (mechanical) modes $\hat{a}_n$ ($\hat{b}_n$) with single-photon radiation pressure coupling $g_{nl}$ are driven by an external laser with frequency $\omega_L^{(n)}$ and amplitude $\Omega_n$. \textbf{b} The interacting optomechanical model described above can be mapped to two decoupled single-mode optomechanical systems with single-cavity mode $\hat{a}_n$ and mechanical oscillator normal mode $\hat{B}_n$.}
	\label{fig:model}
\end{figure}

We consider an optomechanical system composed of two mechanical oscillators, modeled as vibrating
dielectric membranes of identical frequencies, located inside a driven two-mode optical resonator in the single-photon
optomechanical strong regime \cite{chinesreview}, see Fig.~\ref{fig:model}%
(a). The total Hamiltonian in the rotating frame at the frequencies of the
driven lasers is ($\hbar =1$) 
\begin{multline}
\hat{H}=\sum_{n,l=1}^2-\Delta _{n}\hat{a}_{n}^{\dagger }\hat{a}_{n}+\omega
_{m}b_{n}^{\dagger }b_{n}-g_{nl}\hat{a}_{n}^{\dagger }\hat{a}_{n}(\hat{b}%
_{l}^{\dagger }+\hat{b}_{l})\\+\Omega _{n}(\hat{a}_{n}^{\dagger }+\hat{a}_{n}).
\label{H}
\end{multline}
In the above, $\hat{a}_{n}$ ($\hat{b}_{n}$) denotes the $n$th bosonic
annihilation operator of the cavity (mechanical) mode with frequency $\omega_c^{(n)}$ ($\omega_m$). $g_{nl}$
represents the single-photon radiation pressure coupling strength between the $n$th cavity and the $l$th mechanical modes~\cite{gnl1,gnl2,gnl3}. $\Delta
_{n}{=}\omega _{L}^{(n)}{-}\omega _{c}^{(n)}$ accounts for the $n$th pump
detuning between the cavity frequency $\omega _{c}^{(n)}$ and the external
laser field of amplitude $\Omega _{n}$ and frequency $\omega _{L}^{(n)}$. 
As a first step in our protocol, we proceed to diagonalize the Hamiltonian
using the \ transformation%
\begin{eqnarray}
\hat{B}_{1} &=&\hat{b}_{1}\cos (\theta )+\sin (\theta )\hat{b}_{2}, \\
\hat{B}_{2} &=&\hat{b}_{1}\sin (\theta )-\cos (\theta )\hat{b}_{2},
\end{eqnarray}%
where the optomechanical coupling strengths satisfy 
\begin{eqnarray}
g_{12} &=&\tan (\theta )g_{11}=g_{1}, \\
g_{21} &=&-\tan (\theta )g_{22}=g_{2},
\end{eqnarray}
and the set of operators $\{\hat{B}_n\}$ are the so-called mechanical oscillators' normal modes. Note that, recent experimental efforts in a two-membrane setup have tuned the optomechanical couplings $g_1$ and $g_2$ by high-precision positioning of the membranes via piezo control~\cite{Piergentili2018} or through on-chip monolithic arrays of two highly reflective mechanical resonators~\cite{Gartner2018}. The above transformation allows us to rewrite the total Hamiltonian as 
\begin{multline}
\hat{H}=\sum_{n=1}^{2}-\Delta _{n}\hat{a}_{n}^{\dagger }\hat{a}_{n}+\omega
_{m}\hat{B}_{n}^{\dagger }\hat{B}_{n}-g_{n}\hat{a}_{n}^{\dagger }\hat{a%
}_{n}(\hat{B}_{n}^{\dagger }+\hat{B}_{n})\\+\Omega _{n}(\hat{a}_{n}^{\dagger }+%
\hat{a}_{n}).  \label{eq:diagonalized}
\end{multline}%
It is worth emphasizing that, from Eq.~\eqref{eq:diagonalized} the system
can be effectively viewed as a single nonlinear optomechanical unit where
each driven cavity mode is coupled to a single mechanical oscillator normal
mode, see Fig.~\ref{fig:model}(b). To clarify the physical mechanism behind
our proposal, we derive an effective Hamiltonian between the set of normal
modes operator $\{\hat{B}_{n}\}$ and the photonic operator in the
single-photon subspace. To do so, we first switch to a displaced basis that
diagonalizes the radiation pressure interaction via the displacement
operator~\cite{Bose1997,Velazquez2015} 
\begin{equation}
\hat{D}_{n}(\hat{\xi}_{n})=e^{\hat{\xi}_{n}\hat{B}_{n}^{\dagger }-\hat{\xi}%
_{n}^{\dagger }\hat{B}_{n}},\hspace{0.5cm}\hat{\xi}_{n}=\hat{a}_{n}^{\dagger
}\hat{a}_{n}g_{n}/\omega_{m}.
\end{equation}%
Secondly, we perform the unitary transformation 
\begin{equation}
\hat{U}_{n}{=}\mathrm{exp}\left[ i\frac{g_{n}^{2}}{\omega_{m}}(\hat{a}%
_{n}^{\dagger }\hat{a}_{n})^{2}\right] \mathrm{exp}[-i(\omega_{m}\hat{%
B}_{n}^{\dagger }\hat{B}_{n}-\Delta _{n}\hat{a}_{n}^{\dagger }\hat{a}_{n})t],
\end{equation}%
which leads to the displaced Hamiltonian 
\begin{equation}
\hat{H}_{D}{=}\sum_{n=1}^{2}\Omega _{n}e^{-i\Delta _{n}t}e^{-ig\eta _{n}t(2%
\hat{a}_{n}^{\dagger }\hat{a}_{n}+1)}\hat{a}_{n}^{\dagger }\hat{D}^{\dagger
}(\eta _{n}e^{i\omega_{m}t}){+}\mathrm{H.c},  \label{eq:Hd}
\end{equation}%
where $\eta _{n}=g_{n}/\omega_{m}$ is the scaled optomechanical
coupling interaction associated to the $n$th cavity field of the now
decoupled optomechanical cell, see Fig.~\ref{fig:model}(b). By assuming the
resolved-sideband limit where the $n$th cavity linewidth $\kappa _{n}$ is
much smaller than $\omega_{m}$, we can safely choose the
blue-sideband condition, $\Delta _{n}+g_{n}\eta _{n}=\omega_{m}$, in
the single-photon subspace, the so-called photon blockade regime~\cite%
{blockade}, $\kappa _{n}\ll g_{n}\eta _{n}$. The single-photon condition,
allows us to redefine the photonic operators as 
\begin{eqnarray}
\hat{a}_{n} &=&|0\rangle _{nn}\langle 1|=\hat{\sigma}_{-}^{(n)}, \\
\hat{a}_{n}^{\dagger } &=&|1\rangle _{nn}\langle 0|=\hat{\sigma}_{+}^{(n)}.
\end{eqnarray}%
Finally, by invoking the rotating-wave approximation (RWA) in the presence
of weak laser intensity, $\Omega _{n}\ll \omega_{m}$, we can discard
higher frequencies in the dynamics and obtain the following effective
Hamiltonian 
\begin{equation}
\hat{H}_{\mathrm{eff}}=\sum_{n=1}^{2}\hat{\sigma}_{-}^{(n)}\hat{\chi}%
^{(n)}(\eta _{n})\hat{B}_{n}+\hat{\sigma}_{+}^{(n)}\hat{B}_{n}^{\dagger }%
\hat{\chi}^{(n)}(\eta _{n}),  \label{eq:effective}
\end{equation}%
where 
\begin{equation}
\hat{\chi}^{(n)}(\eta _{n})=\eta _{n}\Omega _{n}e^{-\frac{\eta _{n}^{2}}{2}}%
\frac{L_{\hat{B}_{n}^{\dagger }\hat{B}_{n}}^{(1)}(\eta _{n}^{2})}{\hat{B}%
_{n}^{\dagger }\hat{B}_{n}+1},
\end{equation}%
and $L_{n}^{m}(x)$ are associated Laguerre polynomials. The effective
Hamiltonian in Eq.~\eqref{eq:effective} suffices to understand the physical
mechanism behind the generation of a single-phonon Fock state in the
transformed basis $\{\hat{B}_{n}\}$~\cite{Phono_Fock}, as it evidences the
precise value of $\eta _{n}=g_{n}/\omega_{m}$, given by setting $L_{%
\hat{B}_{n}^{\dagger }\hat{B}_{n}}^{(1)}(\eta _{n}^{2})=0$, such that the
dynamics is confined to a sliced subspace of dimension $0\leq \langle \hat{B}%
_{n}^{\dagger }\hat{B}_{n}\rangle =M_{n}$, $M_{n}$ being an integer for the $%
n$th mechanical mode~\cite{Phono_Fock}. Similar to cavity-cooling schemes,
the above proper $\eta _{n}$ adjustment enables to steer the mechanical
degree of freedom towards the normal mode Fock state $|M_{n}\rangle $ by
directing the optical excitations dynamically to such state in the
transformed basis. Notice that the interference effect between displaced
number states and Fock states, which leads to the slicing of the phononic
Hilbert space, has been previously employed in single-mode optomechanical
setups, for instance, to generate mechanical nonclassical states~\cite%
{NooN,PRA.2013.88.063819,PRA.2013.87.053849,Phono_Fock}.

The next step involves to add the driven quantum evolution in the presence
of decoherence channels. To achieve this goal, we use a dressed-state master
equation within the Born-Markov approximation, a suitable representation for
the dissipative dynamics in the strong optomechanical coupling regime~\cite%
{dmaster}, given by 
\begin{multline}
\frac{d\hat{\rho}}{dt}=\sum_{n=1}^{2}-i\left[ \hat{H},\hat{\rho}\right]  \\
+\frac{\kappa _{n}}{2}\mathcal{D}\left[ \hat{a}_{n}\right] \hat{\rho}+\frac{%
\gamma _{n}}{2}(1+\overline{n}_{m}^{(n)})\mathcal{D}\left[ \hat{b}_{n}-\hat{a%
}_{n}^{\dagger }\hat{a}_{n}\eta _{n}\right] \hat{\rho} \\
+\frac{\gamma _{n}}{2}\overline{n}_{m}^{(n)}\mathcal{D}\left[ \hat{b}%
_{n}^{\dagger }-\hat{a}_{n}^{\dagger }\hat{a}_{n}\eta _{n}\right] \hat{\rho}%
+\left( \frac{k_{B}T}{\omega_{m}}\right) 4\eta _{n}^{2}\gamma _{n}%
\mathcal{D}\left[ \hat{a}_{n}^{\dagger }\hat{a}_{n}\right] .
\label{eq:master}
\end{multline}%
In the above dressed-master equation $\mathcal{D}[\hat{O}]:=2\hat{O}\hat{\rho%
}\hat{O}^{\dagger }-\hat{\rho}\hat{O}^{\dagger }\hat{O}-\hat{O}^{\dagger }%
\hat{O}\hat{\rho}$ is the Lindbladian superoperator term, $\kappa
_{n}(\gamma _{n})$ is the photon (phonon) decay (damping) rate, and $%
\overline{n}_{m}^{(n)}(T)=(\exp [\omega_{m}/(k_{B}T)]-1)^{-1}:=%
\overline{n}_{m}^{(n)}$ is the average occupation number at temperature $T$
for a given mechanical frequency $\omega_{m}$ ($k_{B}=1$ throughout
this work). Since $\omega _{c}^{(n)}\gg \omega_{m}$, we have
neglected the photonic mean occupancy number as $\overline{n}_{c}^{(n)}\ll 
\overline{n}_{m}^{(n)}$. It is worth emphasizing that the above master
equation is valid under the assumption that each mechanical mode weakly interacts
with one particular thermal reservoir. However, if these conditions are not satisfied, it is known that in the strong coupling regime, the fact of having a
common thermal reservoir, or effects of non-Markovianity can all lead to an
effective decoupling of some system's normal modes from the
bath~\cite{DFS1,DFS2,3OS}. As a consequence, one needs to modify the above master equation. Those decoupled modes achieved by decoherence-free subspaces~\cite{DFS1,DFS2} or noiseless
subsystems~\cite{3OS} techniques are then quite beneficial for either entirely decouple the normal modes or reducing the decay rates significantly. Techniques that can be addressed in future works.

The main result of our work is to demonstrate that our protocol, within the
condition $\kappa _{n}\gg \gamma _{n}$, takes any initial state of the form 
\begin{equation}
\hat{\rho}(0)=\bigotimes_{n=1}^{2}\hat{\rho}(0)_{c}^{(n)}\otimes \hat{\rho}%
(0)_{m,M_{n}}^{(n)},  \label{eq:initial-state}
\end{equation}%
and drives it asymptotically to a pure decoupled mechanical steady-state in
the normal mode basis as 
\begin{equation}
\hat{\rho}(\infty )_{m}\approx \bigotimes_{n=1}^{2}\left\vert
M_{n}\right\rangle \left\langle M_{n}\right\vert .
\end{equation}%
Notice that in Eq.~\eqref{eq:initial-state}, the initial $n$th mechanical
system in Fock basis is confined within a phonon subspace less or equal than 
$M_{n}$ 
\begin{equation}
\hat{\rho}(0)_{m,M_{n}}^{(n)}=\sum_{l,m=0}^{M_{n}}p_{lm}\left\vert
l\right\rangle \left\langle m\right\vert .
\end{equation}%
Two main features must be clarified: (i) while choosing the initial state
for the photonic subsystems plays no role in the phononic steady-state
generation, without loss of generality, we have employed only vacuum cavity
states throughout our simulations~\cite{QuTIP}. Note that, both theoretical proposals~\cite{Montenegro2018-cooling, Rao2016-cooling, Mancini1998-cooling} and experimental demonstrations~\cite{Schliesser2006-cooling, Teufel2011-cooling, OConnell2010-cooling, Peterson2016-cooling} for cooling the vibrational motion of a mechanical resonator down to its ground state justify the above; (ii) our proposal differs from the usual reservoir engineering schemes since the steady-state in our
work depends on the system's initial state that must not contain any
excitation above $\{M_{n}\}$. To verify our scheme for the generation of mechanical
entanglement, we numerically solve the steady-state of the full master
equation in Eq.~\eqref{eq:master}. We investigate the effects of the
mechanical damping rates and temperature. To evidence that indeed the system
approaches a highly entangled state, and after reversing from the canonical
transformation that diagonalizes the mechanical modes to the uncoupled
basis, i.e., $\{\hat{B}_n\} \rightarrow \{\hat{b}_n\}$, we calculate the
entanglement between the steady-state modes using the negativity~\cite{neg2} 
\begin{equation}
\mathcal{N}(t)=\sum_i \frac{|\varepsilon _{i}|-\varepsilon _{i}}{2},
\end{equation}%
where $\varepsilon _{i}$ are the eigenvalues of the partially transposed
reduced steady-state density matrix. Additionally, to quantify the degree of
mixedness, we compute the complementary purity measure of the composite
mechanical subsystems $\mathcal{P}(t)=\mathrm{Tr}\left[ \hat{\rho}^{2}(t)\right]$. It is worth stressing that our scheme operates in the
nonlinear optomechanical regime, and thus, its bare single-photon
non-Gaussian interaction allows the system to generate non-Gaussian
entangled steady-states for the mechanical oscillators. To highlight the
inherent non-Gaussianity of our optomechanical protocol, we consider as
quantifier the Wigner logarithmic negativity ($\mathrm{WLN}$)~\cite{WLN},
defined by 
\begin{equation}
\mathrm{WLN}=\log\left(\int \left\vert W(q,p)\right\vert dqdp\right),
\label{WLN}
\end{equation}%
where $W(q,p)$ is the Wigner function and $\mathrm{WLN}>0$ when $W(q,p)$ has
a negative part.
\section{Results}\label{sec:results}
Let us commence our analysis by computing the entanglement in the absence of
mechanical damping $\gamma_{n}=0$. In this scenario, given a proper
parameters tuning, the dissipation of the intracavity field is the one
responsible for preparing a pure and highly entangled state between the two
mechanical modes. To obtain the steady-state density matrix, we solve the
full master equation in Eq.~\eqref{eq:master} in the good-cavity regime $%
\langle\hat{\chi}^{(n)}(\eta_{n})\rangle{\sim}\kappa_n$ with the choice of
symmetric parameters 
\begin{equation}
\hat{\chi}^{(1)}(\eta_{1})=\hat{\chi}^{(2)}(\eta_{2})=\kappa_{1}=\kappa_{2}= 10^{-3}\omega_m.
\end{equation}
The above symmetric choice has been made solely for numerical simplicity,
where $\eta_{n}$ depends on the desired normal mode Fock state $|M_n\rangle$
in the transformed basis $\{\hat{B}_n\}$ by setting $L_{\hat{B}_{n}^{\dagger
}\hat{B}_{n}}^{(1)}(\eta _{n}^{2})=0$ in Eq.~\eqref{eq:effective}.
Furthermore, we have considered vacuum states for all the initial bosonic
modes. The choice of the above parameters satisfy both the blue-sideband
condition $(\kappa_{n}\ll\omega_m)$, the single-photon subspace $%
(\kappa_{n}\ll g_{n}\eta_{n})$, and an initial state with no excitations
above the truncated phonon number of the normal mode Fock state. We now have
all the ingredients to calculate the steady-state entanglement between two
mechanical oscillators. 
\begin{figure}[t]
\centering \includegraphics[width=\linewidth]{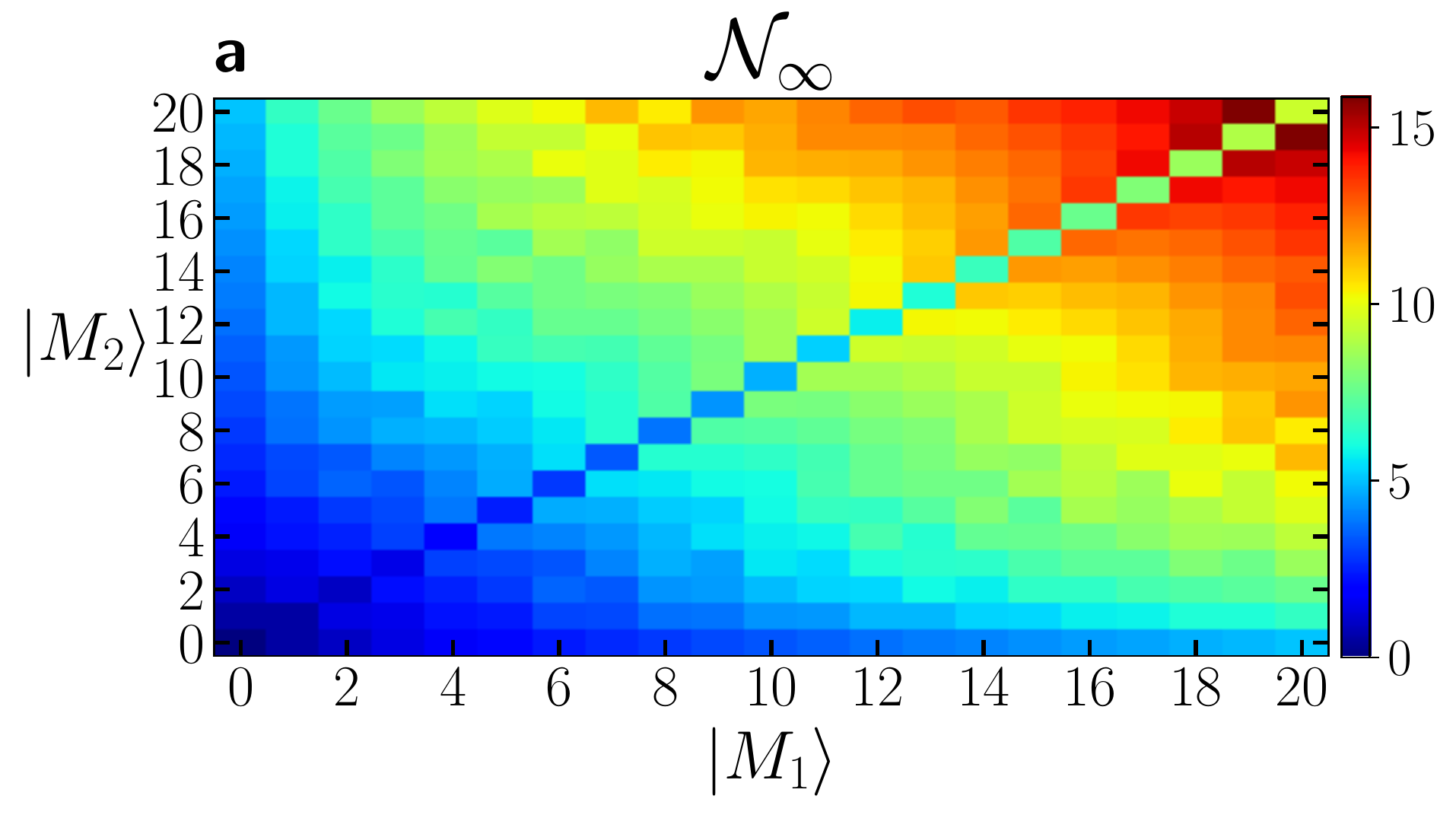} \centering %
\includegraphics[width=\linewidth]{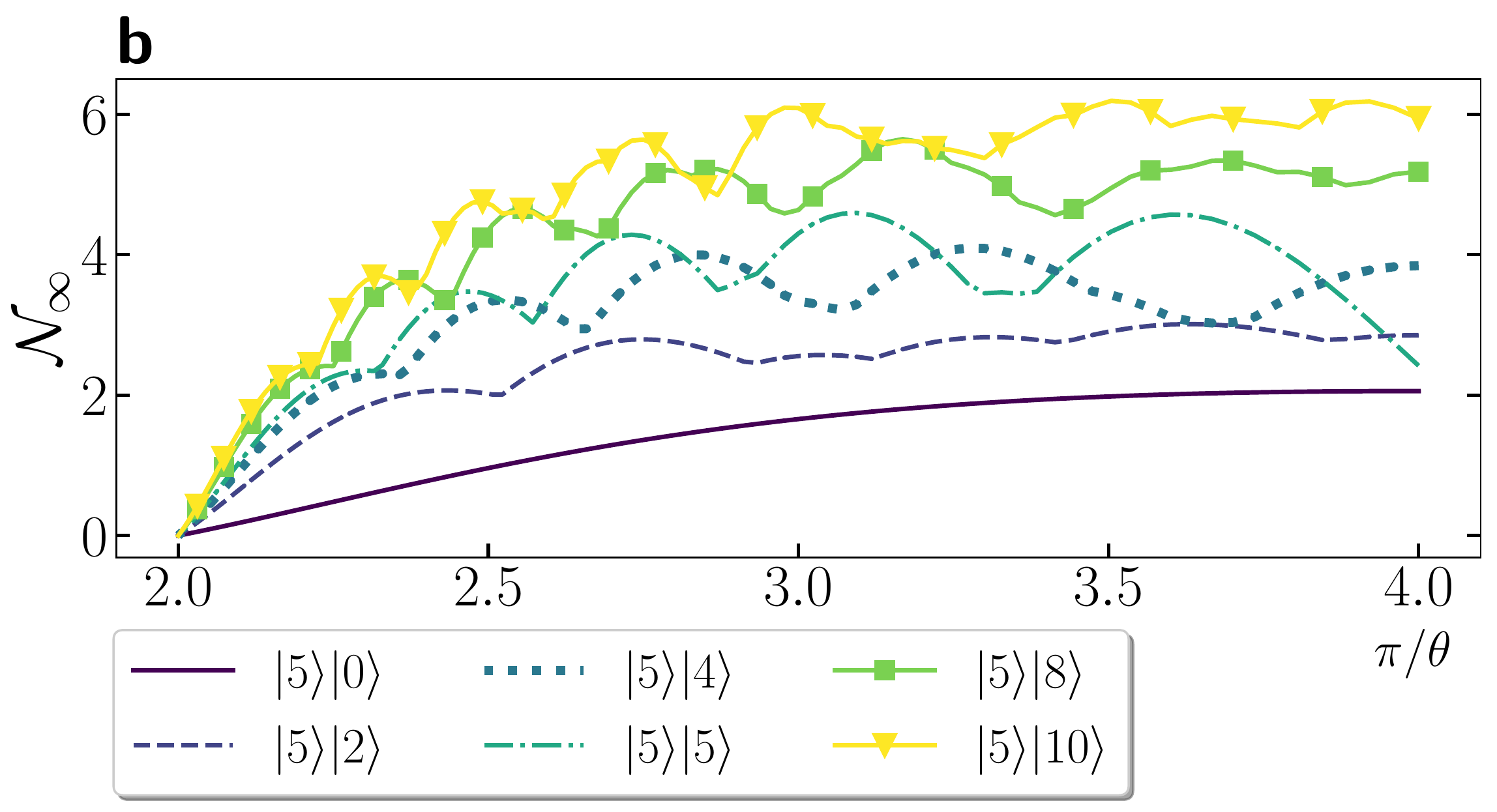}
\caption{\textbf{a} Map between the generated steady-state $\hat{\protect\rho%
}(\infty )=\left\vert M_{1}\right\rangle \left\langle
M_{1}\right\vert\otimes \left\vert M_{2}\right\rangle \left\langle
M_{2}\right\vert$ in the normal mode Fock state and its corresponding two
oscillator steady-state negativity $\mathcal{N}_\infty$ for $\protect\theta{=%
}\protect\pi/4$, i.e., $g_{12}=g_{11}=g_{1}$ and $g_{21}=-g_{22}=g_{2}$. The negativity is computed upon reversing the Bogoliubov
transformation from $\{\hat{B}_n\}$ to the uncoupled basis $\{\hat{b}_n\}$. 
\textbf{b} The steady-state negativity $\mathcal{N}_{\infty}$ as a function
of $\protect\pi/\protect\theta$ for several target normal mode Fock states $%
|M_{1}\rangle\otimes|M_{2}\rangle$. Different rotation angle $\protect\theta$
entails different two-mode oscillators entanglement. For both figures $\hat{%
\protect\chi}^{(1)}(\protect\eta_{1}){=}\hat{\protect\chi}^{(2)}(\protect\eta%
_{2}){=}\protect\kappa_{1}{=}\protect\kappa_{2}{=}10^{-3}\omega_m$, $\protect\gamma_n{=}0$.}
\label{fig:fig2}
\end{figure}
\begin{figure}[t]
\centering \includegraphics[width=\linewidth]{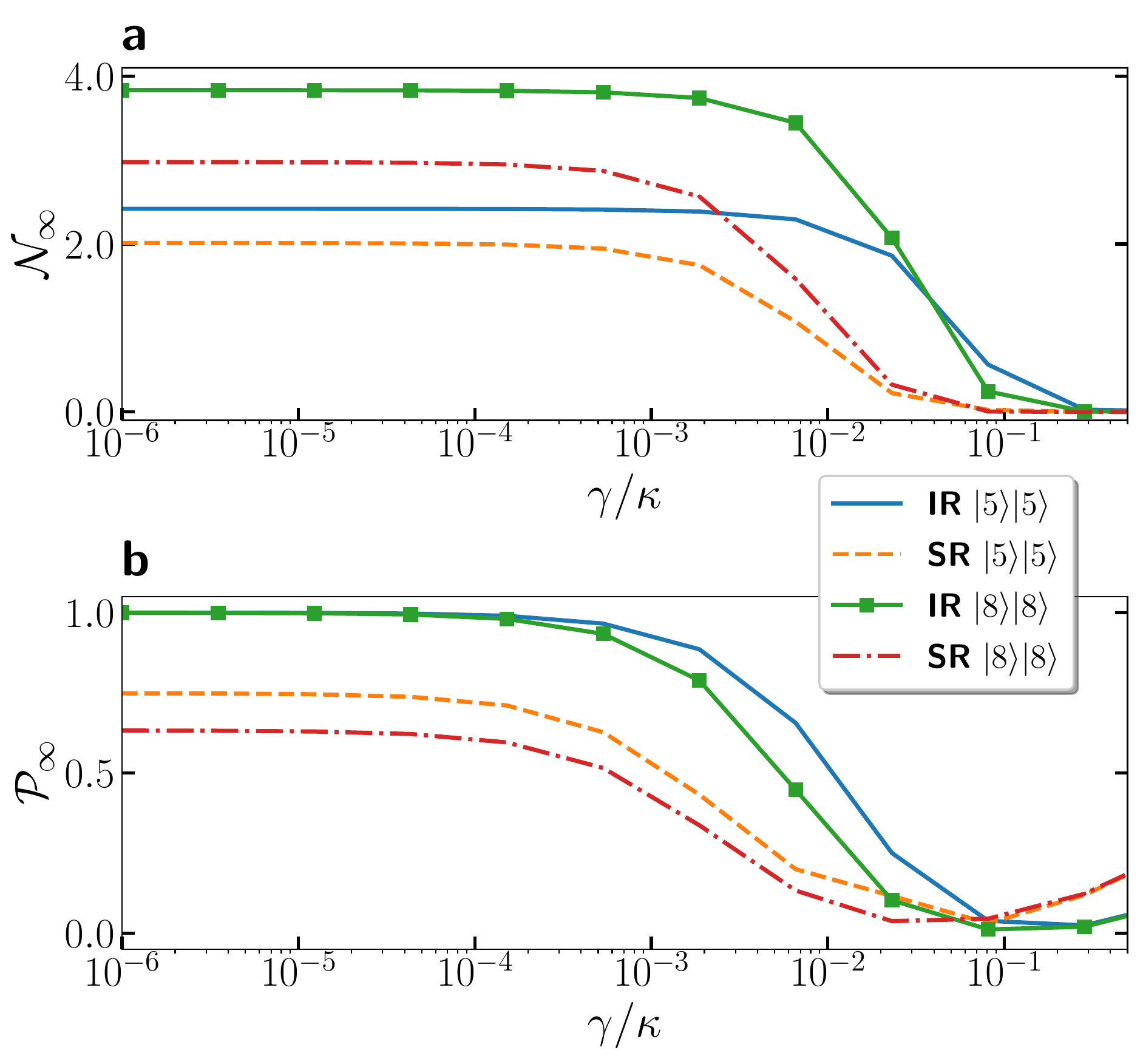}
\caption{\textbf{a} Steady-state entanglement $\mathcal{N}_{\infty}$ and 
\textbf{b} steady-state purity $\mathcal{P}_{\infty}$ as a function of $%
\protect\gamma/\protect\kappa$ for the normal mode Fock states $%
|5\rangle|5\rangle$ and $|8\rangle|8\rangle$. The solid lines account for
the case of each mechanical normal mode coupled to an individual reservoir
(IR), whereas the dashed lines are associated to a single shared (SR)
photonic reservoir for both mechanical normal modes. Other values are $\hat{%
\protect\chi}^{(1)}(\protect\eta_{1}){=}\hat{\protect\chi}^{(2)}(\protect\eta%
_{2}){=}\protect\kappa_{1}{=}\protect\kappa_{2}{=}10^{-3}\omega_m$ and zero temperature.}
\label{fig:fig3}
\end{figure}

In Fig.~\ref{fig:fig2}(a), we show the map between the normal mode
Fock states $|M_1\rangle$ and $|M_2\rangle$ of the enginereed steady-state
density matrix $\hat{\rho}(\infty)=\left\vert M_{1}\right\rangle
\left\langle M_{1}\right\vert\otimes \left\vert M_{2}\right\rangle
\left\langle M_{2}\right\vert$ and its corresponding steady-state negativity 
$\mathcal{N}_\infty$ for $\theta{=}\pi/4$, i.e., $g_{12}=g_{11}=g_{1}$ and $g_{21}=-g_{22}=g_{2}$. As seen from the figure, for the choice of $%
|M_1\rangle=|M_2\rangle$ the steady-state entanglement increases
monotonically as we keep generating higher Fock states equally in both
mechanical modes (the diagonal curve in the heatmap). We remind the reader
that the generation of a normal mode Fock state $|M_n\rangle$ depends only
of the choice of $\eta_n=g_n/\omega_m$, i.e., $L_{M_n}^{(1)}(\eta_n)=0$%
. On the other hand, for a fixed normal mode Fock state $|M_1\rangle$ such
that $|M_1\rangle>|M_2\rangle$ the steady-state entanglement increases as $%
|M_2\rangle$ increases with an evident dip when the normal mode Fock state $%
|M_2\rangle$ matches $|M_1\rangle$. As $|M_2\rangle$ becomes larger than $%
|M_1\rangle$, the steady-state negativity continue increasing as $|M_2\rangle
$ increases. The dip can be understood due to the fixed choice of $%
\theta=\pi/4$ [see the upcoming discussion in Fig.~\ref{fig:fig2}(b)]. The
above description can be summarized as following: given any set of
engineered normal mode Fock states $|M_1\rangle$ and $|M_2\rangle$, the
steady-state entanglement can be increased by generating a higher normal
mode Fock state $|M_1\rangle$, $|M_2\rangle$, or both simultaneously.
Strikingly, our protocol proves to generate stronger entanglement than the
one obtained by a direct two-mode squeezing coupling in the same regime of
parameters. This is because the upperbound $\mathcal{N}_{\infty}\leq 2^{\ln
(2)}$ arising from stability conditions in the latter two-mode squeezing
coupling scheme~\cite{stable}. Such upperbound entanglement \textit{handicap}%
, which is shared among linearized optomechanical Hamiltonians within the
blue-sideband condition, is lifted in our protocol thanks to the inclusion
of the inherent nonlinear (non-Gaussian) optomechanical interaction.  

We explore the
rotation angle $\theta$ from the Bogoliubov transformation which is a
function of the optomechanical coupling asymmetry. In Fig.~\ref{fig:fig2}(b), we
show the steady-state entanglement $\mathcal{N}_{\infty}$ as a function of
the angle $\theta \in (\pi/2, \pi/4]$ for different normal mode Fock states $%
|M_{1}\rangle$ and $|M_{2}\rangle$. Notably, in general, the maximum
entanglement is achieved for unequal optomechanical couplings $g_{nl} $, and the optimal angle $\theta$ that
maximizes $\mathcal{N}_{\infty}$ depends on the choice of the target normal
mode Fock state. Notice that, we have explicitly included the case of the
entanglement as a function of $\theta$ for the target states $%
|5\rangle\otimes|5\rangle$. For this scenario, it is evident that the choice
of $\theta=\pi/4$ highly deteriorates the entanglement between mechanical
oscillators which corresponds to the noticeable dip observed in Fig.~%
\ref{fig:fig2}(a). 
\begin{figure}[t]
\centering \includegraphics[width=1\linewidth]{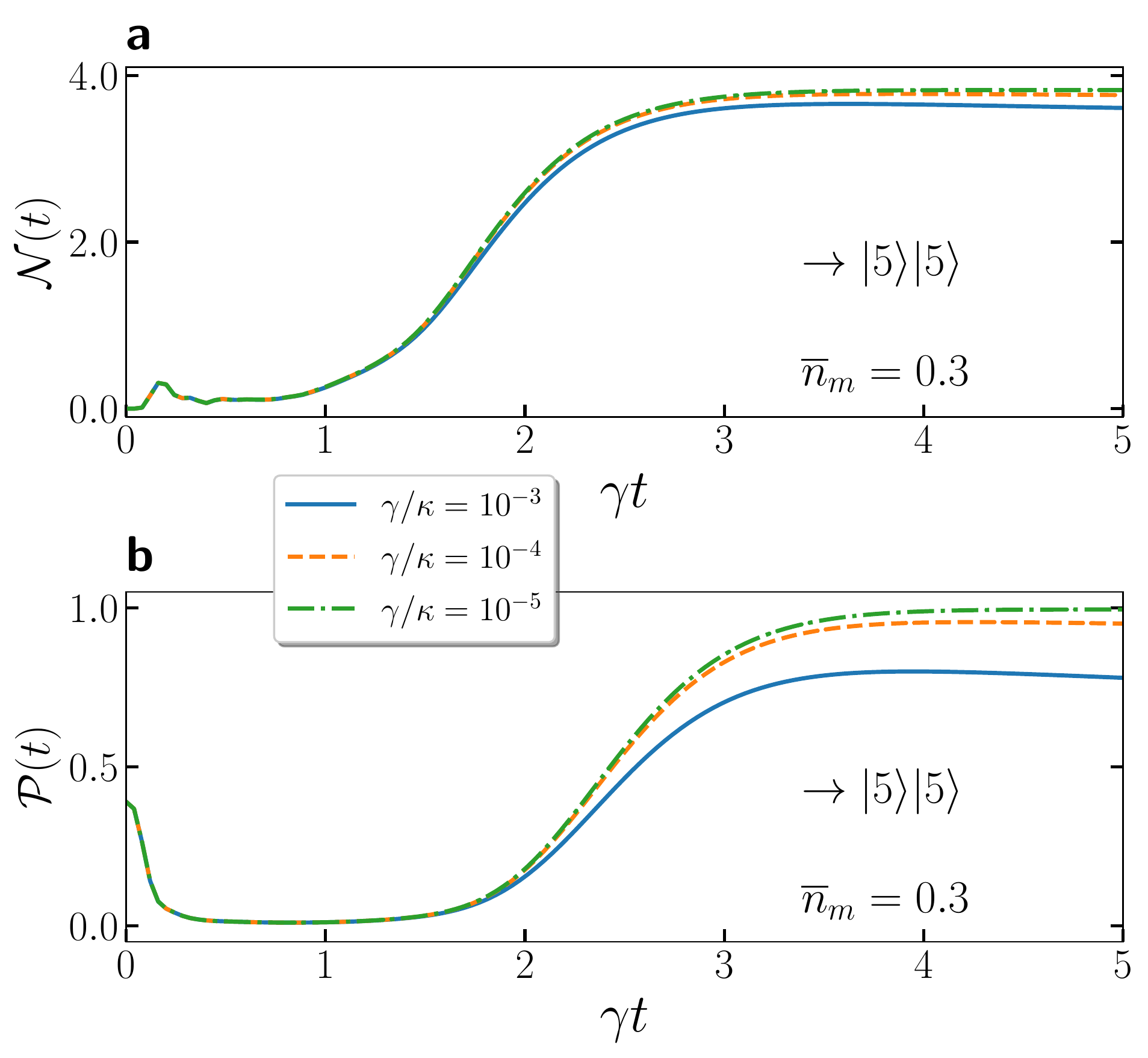}
\caption{\textbf{a} Entanglement dynamics $\mathcal{N}(t)$ and \textbf{b}
purity dynamics $\mathcal{P}(t)$ as a function of the scaled time $\protect%
\gamma t$ for three different decoherence ratios $\protect\gamma/\protect%
\kappa$. We aim to reach the normal mode Fock states $|5\rangle |5\rangle$
by considering both mechanical thermal reservoirs at the same temperatures $%
\overline{n}_{m}^{(1)}=\overline{n}_{m}^{(2)}=\overline{n}_m=0.3$.}
\label{fig:fig4}
\end{figure}

While the previous analysis was performed in the absence of mechanical
damping $\gamma_{n}=0$, one key point is to study up to which values of the
relevant decoherence parameters our protocol can be accomodated. To
quantitatively address the influence of mechanical decoherence, we consider
in the following a finite mechanical damping $\gamma_n=\gamma$ for two
scenarios, namely:
\begin{itemize}
    \item The case of a single shared photonic reservoir (SR),
where the two mechanical modes are coupled to a single cavity mode, i.e., $\hat{a}^\dagger\hat{a}(\hat{B}_1+\hat{B}_2 + \text{H.c})$,
\item and the case in which each mechanical normal mode couples to an individual
reservoir (IR).
\end{itemize}

In Fig.~\ref{fig:fig3}(a), we compute the steady-state
negativity $\mathcal{N}_\infty$ as a function of the decoherence ratio $%
\gamma/\kappa$ for two normal mode Fock states $|5\rangle|5\rangle $ and $%
|8\rangle|8\rangle$. As the figure shows, for a given two-mode target state,
a shared reservoir results always in a detrimental effect for the
steady-state entanglement. In Fig.~\ref{fig:fig3}(b), we present the
steady-state purity $\mathcal{P}_\infty$ as a function of the decoherence
ratio $\gamma/\kappa$ for the same target states. As evident from the
figure, the IR case not only provides higher entanglement but also delivers
higher purity for the same range of decoherence ratio $\sim 10^{-3}$ to $%
\sim 10^{-2}$. Interestingly, for the SR scenario, the steady-state purity
considerably decreases for the same range of decoherence parameters. One can thus assert that the production of highly-pure-entangled states are robust up
to $\gamma/\kappa \sim 10^{-2} (10^{-3})$ for the IR (SR) scenario. While it
is clear that SR leads to weaker entanglement and greater mixture than IR,
our protocol still remains valid for the generation of phononic bipartite
entanglement. 
\begin{figure}[t]
\centering \includegraphics[width=\linewidth]{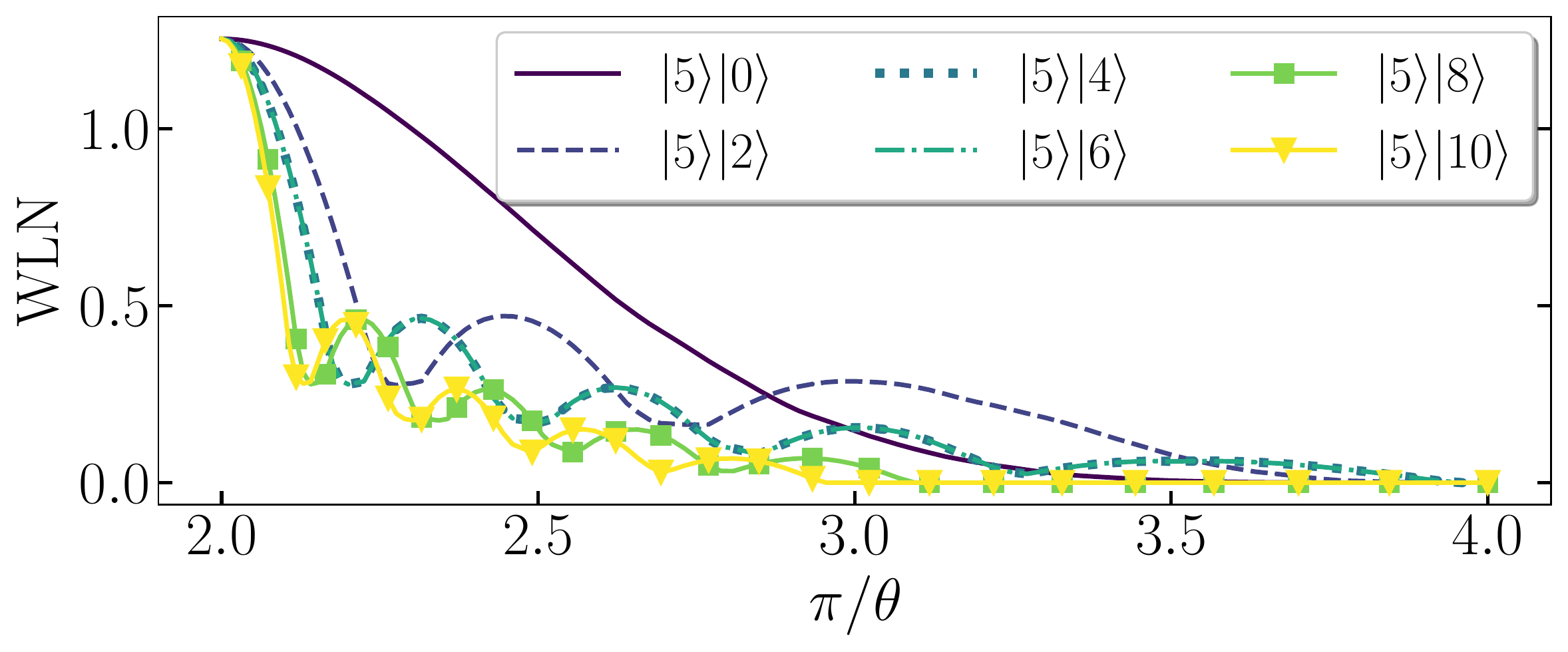}
\caption{Non-Gaussianity of the entangled state associated to the normal
mode Fock states $|M_1\rangle|M_2\rangle$ quantified by the Wigner
logarithmic negativity $(\mathrm{WLN})$. The figure shows the WLN computed for the first mechanical mode as a function of the angle $\protect\theta$.}
\label{fig:fig5}
\end{figure}

We now study the effects of the mechanical reservoir temperature in the
dynamics to reaching the desired mechanical oscillators' steady-state. To do
so, we fix the thermal occupation of the mechanical baths to
\begin{equation}
\overline{n}^{(1)}_m=\overline{n}^{(2)}_m=\overline{n}_m=0.3
\end{equation}
for a given target state
while varying the decoherence ratio $\gamma/\kappa$. Notice that, thermal
occupation of $\overline{n}_m\sim 0.4$ can already be achieved experimentally in
mechanical resonators operating in the microwave regime at millikelvin
temperatures~\cite{temp}. Let us first investigate the impact of the
mechanical reservoir temperature on the entanglement dynamics $\mathcal{N}(t)
$ for reaching the specific normal mode Fock states $|5\rangle |5\rangle$.
In Fig.~\ref{fig:fig4}(a), we plot $\mathcal{N}(t)$ as a function of the
scaled time $\gamma t$ for different decoherence ratios $10^{-5} \leq
\gamma/\kappa \leq 10^{-3}$ aiming $|5\rangle |5\rangle$. As the figure
shows, for low mechanical reservoir temperatures the entanglement dynamics is barely affected for a wide range of decoherence parameters. To
further investigate the impact of the temperature, in Fig.~\ref{fig:fig4}(b)
we show the purity dynamics $\mathcal{P}(t)$ as a function of the scaled
time $\gamma t$ for the same decoherence ratios. Notably, while the purity
is less robust as compared to the entanglement strength, we stress that
higher temperatures will lead to unstable regimes of the steady-state.
However, the scheme remains efficient in the timescale $\tau$ that obeys
\begin{equation}
    (\kappa_{n})^{-1}\ll\tau\ll(\overline{n}^{n}_m\gamma _{n})^{-1},
\end{equation}
and consequently, our proposal arises as a useful scheme for the generation of
resourceful quantum states in the presence of non-zero mechanical reservoir
temperature. 

Finally, the non-Gaussianity of the steady-state is
investigated in Fig.~\ref{fig:fig5}, where we have plotted the $\mathrm{WLN}$ as a function of
the angle $\theta$ for the same set of parameters of Fig.~\ref{fig:fig2}. As evident from the figure, the WLN reaches the maximum value at the \textit{unphysical} value of $\theta\rightarrow\pi /2$ ($g_{nl}\rightarrow \infty $) regardless the target normal modes. In such scenario, the system is in a product state and target states $|M_{1}\rangle|M_{2}\rangle$ showing higher entanglement [see Fig.~\ref{fig:fig2}(b)] tends to exhibit lower $\mathrm{WLN}$ for unequal couplings. Interestingly, in the range of $\pi/2 < \theta \leq \pi/4$ a wide non-zero region of non-Gaussianity can be attained with our dissipative optomechanical scheme. Indeed, in addition to the strongly nonlinear interaction regime, the main condition to obtaining non-Gaussian states is the arising high asymmetry between the optomechanical coupling strengths~\cite{gnl1,gnl2,gnl3}. Finally, one can notice that, for the same phononic mean excitation number~\cite{WLN}, the amount of non-Gaussianity allowed for our scheme proves to be higher than the one obtained by means of photon subtraction/addition of Gaussian states~\cite{add1,add2}.

\section{Experimental Feasibility}\label{sec:experimental-feasibility}
Our theoretical proposal for achieving dissipative multimode mechanical entanglement poses nontrivial experimental challenges, mainly involving the strength and tunability of the required optomechanical couplings. In what follows, we briefly present how these difficulties have been addressed in state-of-the-art cavity optomechanics, and hence, the promising future of our protocol's feasibility in practice.

\subsection{On the proper adjustments of the single-photon optomechanical couplings}
Let us first start the discussion on the needed optomechanical tunability of the coupling strengths
\begin{eqnarray}
\nonumber g_{12} &=&\tan (\theta )g_{11}=g_{1}, \\
g_{21} &=&-\tan (\theta )g_{22}=g_{2}, \label{new_gs}
\end{eqnarray}
which leads to the diagonalization of the Hamiltonian in Eq.~\eqref{eq:diagonalized} written in their mechanical normal modes. Indeed, our protocol demands to make each normal mode $\hat{B}_n$ simultaneously bright for its corresponding optical driving mode. Although quite challenging, recent experimental efforts in optomechanical arrays have been carried out in this direction~\cite{Piergentili2018, Gartner2018, Nair2017, Weaver2017}. For instance, unlike the high-precision positioning of the mechanical elements for tuning the optomechanical couplings, similar coupling strengths have been achieved, avoiding extremely precise positioning through on-chip monolithic arrays of two highly reflective mechanical resonators~\cite{Gartner2018}. In such a setup, almost identical mechanical oscillators with frequencies of $\sim 150$ kHz and quality factors of the order of $10^6$ can be adequately adjusted such that the center-of-mass coupling of the mechanical array vanishes. In such a case, although not experimentally observed for multimode optomechanical arrays, it has even been suggested that the strong single-photon optomechanical regime can be achievable $g_0>\omega_m$~\cite{Xuereb2012}. This can be understood as the cavity field becomes resonant with the inner cavity modes and thus couples strongly to the relative motion of the membranes, i.e., constructive interference inside the resonator's internal volume. Hence, further enhancement of the optomechanical coupling can even be improved by decreasing the separation between membranes~\cite{Gartner2018}. 

Perhaps the closest experimental setup to our theoretical protocol has been recently achieved~\cite{Piergentili2018}. Indeed, the first two membrane setup has been realized with commercially available square thin layer membranes of low-tensile stress silicon nitride (SiN) and high-tensile stress Si$_3$N$_4$ with millimeter sides and nanometric thicknesses. The membranes setup, mounted on piezoelectrical controls, are dimensionally, optically, and mechanically characterized, showing two remarkable outcomes: (i) the apparent enhancement of the optomechanical coupling with respect to the single membrane case due to constructive interference~\cite{Xuereb2012} with an optomechanical gain of $\sim 2.74$; and (ii) the experimental on-demand tunability of the optomechanical coupling as in Eq.~\eqref{new_gs}, where it is even possible to couple only one membrane to the cavity field while decoupling the other one and simultaneously making both mechanical normal modes bright for an optical mode by shifting the membrane positions via high-precision piezo control. 

\subsection{On the strength of the single-photon optomechanical couplings}
The above experimental progress in optomechanical arrays, including the first reported calibration and fully characterization of two membranes setup in practice~\cite{Piergentili2018}, still show weak $g\ll \omega_m$ single-photon optomechanical couplings. Thus, still far away from the requirements of our theoretical proposal to succeed. For instance, $g_{1}=2\pi \times 0.30 \text{Hz}$ and $g_{2}=2\pi \times 0.28 \text{Hz}$ for the low-stress and high-stress membranes with frequencies of $\sim 10^2 \text{kHz}$ have been reported~\cite{Piergentili2018}, respectively. To determine the needed scaled optomechanical coupling to generate a normal single-mode Fock state $|M\rangle$, we recall that one needs to satisfy 
\begin{equation}
\hat{\chi}^{(n)}(\eta _{n}) \sim L_{\hat{B}_{n}^{\dagger }\hat{B}_{n}}^{(1)}(\eta _{n}^{2}) = 0.\label{eq:laguere}
\end{equation}
\begin{figure}[t]
\centering \includegraphics[width=1\linewidth]{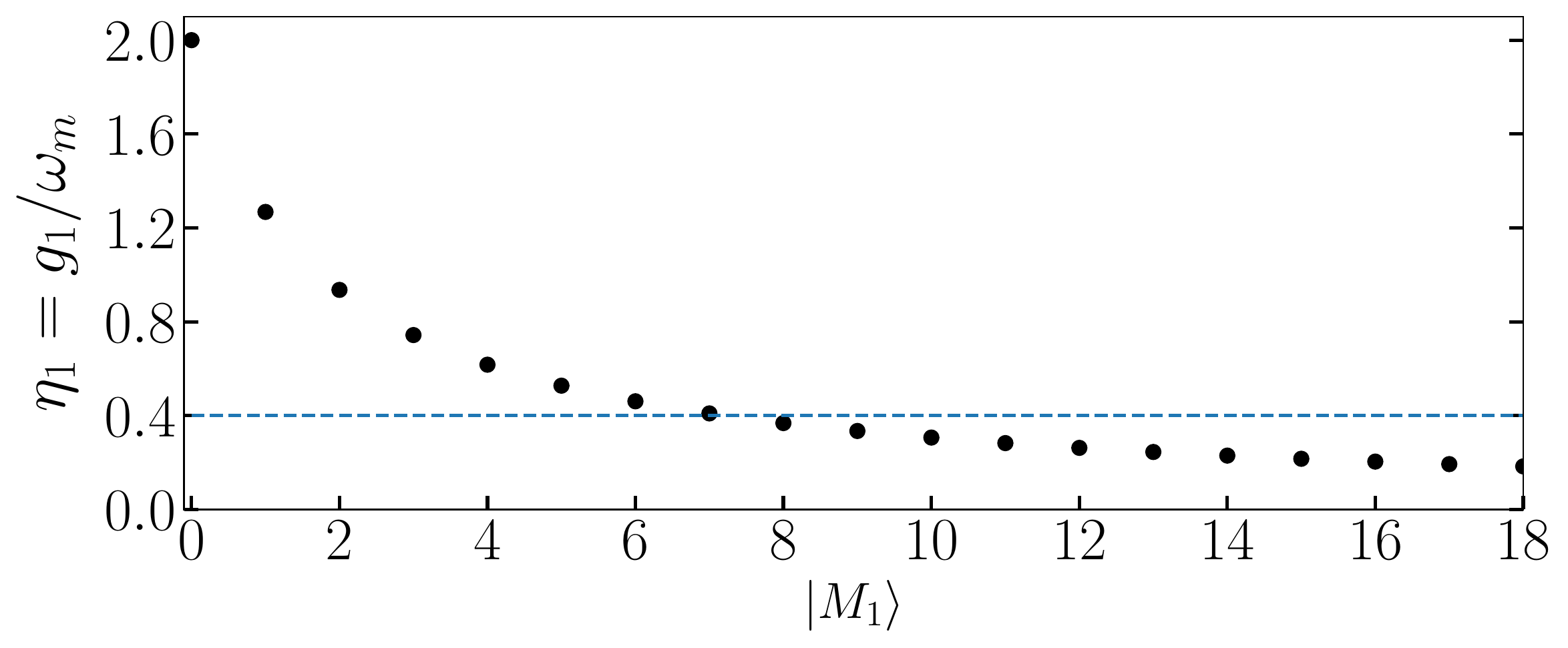}
\caption{Required optomechanical coupling $\eta_1 = g_1/\omega_m$ to satisfy $L^{(1)}_M(\eta_1^2)=0$ as a function of the normal single-mode Fock number state $|M_1\rangle$.}
\label{fig:laguerre}
\end{figure}
Note that the proper adjustments for the above can be attained, in principle, by placing the mechanical objects with the aid of piezo controls~\cite{Nair2017} or using an on-chip monolithic technique~\cite{Gartner2018}. In Fig.~\ref{fig:laguerre}, we plot the required optomechanical coupling $\eta_1=g_1/\omega_m$ to satisfy Eq.~\eqref{eq:laguere} as a function of the Fock number state $|M_1\rangle$. As the figure shows, to generate high bi-modal phonon number states $|M_1\rangle|M_2\rangle$ the scaled couplings $\eta_n=g_n/\omega_m$ should be in the order of $(\eta_1, \eta_2) \lesssim 0.4$. Also notice that \textit{failing} in tuning the associate Laguerre expression to zero would entail the production of, yet not deterministically, another adjacent normal single-mode Fock number state $|M_1\rangle$ (for a more detailed discussion, see Ref.~\cite{Phono_Fock}). Together with the resolved sideband and photon blockade requirements, our protocol primarily demands
\begin{equation}
    \omega_m \gtrsim g_n \gg (\kappa_n, \gamma_n).
\end{equation}
Since the strong coupling regime has not been experimentally achieved in multimode cavity optomechanics, it is a complex endeavor to justify our theoretical findings in such a regime. Nonetheless, without falling into direct extrapolation, we could mention some remarkable experiments towards the strong single-photon optomechanical regime involving single-mode interactions. Indeed, optomechanical systems in the classical regime (with a suggested quantum level, see Supplemental Material in Ref.~\cite{Shkarin2014}), can achieve the strong optomechanical regime in the unresolved sideband. The two-mode membrane interacting with a single-mode cavity field shows a mechanical mode $\omega_m(3,3) = 2\pi \times 5.092 \text{MHz}$ coupled to the cavity field with $g_0(3,3) = 2\pi \times 1.05 \text{MHz}$, i.e., a scaled optomechanical value of $\eta \sim 1/5$. While the conventional manner to enhance the single-photon interactions involve a strong driving of the cavity~\cite{Optomechrev, Teufel2011, Groblacher2009}, leading to an interaction of two harmonic oscillators, there are other ways to enhance such an interaction. In particular, we could mention the entirely detachment of the mechanical oscillator employing levitated silica nanoparticles at room temperature~\cite{Sommer2021}, where the transition to the strong coupling regime entails optomechanical frequencies $g\sim 10 \text{kHz}$ at moderate resolved sideband $g > \kappa/4$ and mechanical frequency $\omega_m\sim 10^2 \text{kHz}$, i.e., $\eta \sim 0.1$. A novel approach includes the enhancement of the optomechanical coupling via hybridizing the system with a qubit system~\cite{Neumeier2018, Aporvari2021}. There, the inherent nonlinearity of the two-level system component makes this setup very suitable for achieving strong optomechanical interaction at the single-photon level~\cite{Aporvari2021}.

\section{Concluding remarks}\label{sec:conclusions}

This work proposes a theoretical protocol for generating steady-state non-Gaussian mechanical entanglement in the strong single-photon optomechanical regime. Unlike the dominant proposals where the optomechanical interaction is linearized, we exploit the inherent non-Gaussian light-matter coupling. Our system is composed of two mechanical oscillators, where each mechanical mode couples nonlinearly to two-modes laser-driven cavity fields. We analytically show that for an appropriate laser-driven and tunning of the optomechanical coupling interaction, one can steer the mechanical oscillators' normal modes asymptotically towards bipartite Fock states in the normal mode basis. Notably, reversing to the uncoupled mechanical basis shows that it is possible to achieve a high degree of steady-state entanglement. In addition, we study the effects of decoherence and thermal fluctuations. Even in these cases, our scheme shows to be quite robust under such detrimental effects. Finally, we study the conditions to obtain non-Gaussianity of the generated mechanical steady-state, showing that the central condition for this to happen is the arising asymmetry between optomechanical couplings. Strikingly, our dissipative optomechanical scheme generates a higher amount of non-Gaussianity than the paradigmatic examples of photon subtraction/addition schemes for the same end.

\section*{Acknowledgments}
V.M. thanks the National Natural Science Foundation of China (Grant No. 12050410251), the Chinese Postdoctoral Science Fund (Grant No. 2018M643435), and the Ministry of Science and Technology of China (Grant No. QNJ2021167004)

\end{document}